\documentclass{iau}

\usepackage{amsmath}
\usepackage{graphicx}
\usepackage{multirow}

\begin{document}

\lefttitle{Luca Tortorelli et al.}
\righttitle{ProMage: fast galaxy magnitudes emulation}

\jnlPage{1}{7}
\jnlDoiYr{2025}
\doival{10.1017/xxxxx}

\aopheadtitle{Proceedings IAU Symposium}

\title{\textsc{ProMage}: fast galaxy magnitudes emulation combining SED forward-modelling and machine learning}

\author{Luca Tortorelli$^{1}$,  Silvan Fischbacher$^{2}$,  Aaron S.G. Robotham$^{3}$,  C\'eline Nussbaumer$^{2}$,  Alexandre Refregier$^{2}$}
\affiliation{$^{1}$Universit\"ats-Sternwarte, Fakult\"at f\"ur Physik, Ludwig-Maximilians-Universit\"at M\"unchen, Scheinerstr. 1, 81679 M\"unchen, Germany, \email{luca.tortorelli@physik.lmu.de}.}
\affiliation{$^{2}$Institute for Particle Physics and Astrophysics, ETH Zurich, Wolfgang-Pauli-Strasse 27, CH-8093 Zurich, Switzerland}.
\affiliation{$^{3}$ICRAR, The University of Western Australia, 7 Fairway, Crawley WA 6009, Australia.}

\begin{abstract}
We present \textsc{ProMage}, a feed-forward neural network that emulates the computation of observer- and rest-frame magnitudes from the generative galaxy SED package \textsc{ProSpect}. The network predicts magnitudes conditioned on input galaxy physical properties, including redshift, star formation history,  gas and dust parameters. \textsc{ProMage} accelerates magnitude computation by a factor of $10^4$ compared to \textsc{ProSpect}, while achieving per-mille relative accuracy for $99\%$ of sources in the test set across the $g,r,i,z,y$ Hyper Suprime-Cam bands. This acceleration is key to enabling fast inference of galaxy physical properties in next-generation Stage IV surveys and to generating large catalogue realisations in forward-modelling frameworks such as \textsc{GalSBI-SPS}.
\end{abstract}


\maketitle

\section{Introduction}

We are entering an era of unprecedented astronomical datasets with the advent of Stage IV galaxy surveys \citep{Albrecht2006}, such as Euclid \citep{Mellier2024}, the Vera C. Rubin Observatory’s Legacy Survey of Space and Time (hereafter,  Rubin-LSST,    \citealt{Ivezic2019}), DESI \citep{DESI2022}, and 4MOST \citep{Dejong2019}. These surveys will measure positions,  magnitudes,  and shapes for billions of galaxies, and redshifts for tens of millions of them,  enabling transformative advances in cosmology and galaxy evolution. However, the statistical power of these datasets makes systematic uncertainties the dominant limitation. For example, weak-lensing cosmology depends critically on accurate galaxy redshift distributions \citep{NewmanGruen2022}, motivating the use of forward-modelling approaches that generate realistic synthetic surveys constrained against real data to provide the required accuracy on the redshift distributions \citep{Fischbacher2025,Tortorelli2025,Thorp2025}. At the same time, the inference of galaxy physical properties for billions of galaxies demands orders-of-magnitude improvements in modelling efficiency.

Both forward-modelling and the inference of physical properties rely on spectral energy distribution (SED) generative codes, which connect galaxy physical properties to their observable stellar emission through stellar population synthesis (SPS; \citealt{Conroy2013}). Codes such as \textsc{FSPS} \citep{Conroy2009} and \textsc{ProSpect} \citep{Robotham2020} provide physically consistent predictions, but their runtimes (tens of seconds for \textsc{FSPS}, tens of milliseconds for \textsc{ProSpect}) remain prohibitive for generating many synthetic survey realisations or running large-scale Monte Carlo Markov Chain (MCMC) inference of galaxy properties. Accelerating SPS predictions while preserving accuracy is therefore essential (see also \citealt{Hearin2023,Alsing2020}).

To address this, we introduce \textsc{ProMage} (\textsc{\textbf{Pro}Spect} \textbf{Mag}nitude \textbf{e}mulator), a feed-forward neural network that emulates galaxy magnitudes in observer and rest frames computed with \textsc{ProSpect}.  Trained on physical inputs (redshift, star formation history, dust, gas parameters), \textsc{ProMage} predicts individual bands independently, simplifying extension to new filters and parameter spaces. It delivers a factor of $\sim10^4$ speed-up in computation with respect to \textsc{ProSpect}, evaluating magnitudes for $10^5$ sources in less than half a second, while maintaining per-mille accuracy across Hyper-Suprime Cam (HSC) $g,r,i,z,y$ bands. This performance enables both efficient forward-modelling with \textsc{GalSBI-SPS} \citep{Tortorelli2025} and scalable inference of galaxy properties for Stage IV surveys, including amortised simulation-based inference \citep{Cranmer2020} and accelerated MCMC. \textsc{ProMage} is already implemented within \textsc{GalSBI-SPS} and will also be released as a standalone module for \textsc{ProSpect}.

\begin{table}
 \centering
 \caption{Prior range for the network input parameters}\label{prior-table}
 {\begin{tabular}{@{\extracolsep{\fill}}lcr}
    \midrule
    Parameter Name& Logarithmic&
     Prior range\\
    \midrule
    z& No &   $[0,5]$\\
    $\mathrm{mSFR}$&         Yes&  $[-3,4]$\\
    $\mathrm{mpeak}$ &         No&   $[-(2+t_{\mathrm{lb}}), 13.4 - t_{\mathrm{lb}}]$\\
    $\mathrm{mperiod}$ &         Yes&   $[\log{(0.3)}, 2]$\\
    $\mathrm{mskew}$ &         No&   $[-0.5,1]$\\
    $\mathrm{Z_{final}}$ &         Yes&   $[-4,-1.3]$\\
    $\log{U}$ &     No &   $[-4,-1]$\\
    $\tau_\mathrm{birth}$&         Yes&   $[-2.5,1.5]$\\
    $\tau_\mathrm{screen}$ &         Yes&   $[-2.5,1]$\\
    $\alpha_\mathrm{SF, birth}$ &         No&   $[0,4]$\\
    $\alpha_\mathrm{SF,,screen}$ &         No&   $[0,4]$\\

    \midrule
    \end{tabular}}

\end{table}

\section{Training sample}

We train \textsc{ProMage} on observer- and rest-frame magnitudes generated with \textsc{ProSpect} in the $g,r,i,z,y$ HSC bands. HSC is an ideal test case, given its depth and galaxy density, making it a precursor to Stage IV surveys. Tests conducted on different neural network architectures show that prediction accuracy improves, at fixed training set size, when networks are trained per band, consistent with results from \cite{Alsing2020,Thorp2025}. 

\textsc{ProMage} predicts total galaxy magnitudes conditioned on the physical inputs used by \textsc{ProSpect}, excluding AGN for now. Stellar emission is based on a custom single stellar population (SSP) created with \textsc{ProGeny} \citep{Robotham2025,Bellstedt2025},   using \textsc{MIST} isochrones \citep{Dotter2016}, C3K spectra \citep{Conroy2018}, and a \cite{Chabrier2003} IMF.  Star-formation histories follow the truncated skewed Normal form of \cite{Robotham2020}, parameterised by the peak SFR ($\mathrm{mSFR}$), peak time ($\mathrm{mpeak}$), width ($\mathrm{mperiod}$), and skewness ($\mathrm{mskew}$). Gas-phase metallicity histories are tied to stellar mass growth \citep{Driver2013,Bellstedt2021}, governed by the final metallicity $\mathrm{Z_{final}}$. Coupled with the ionisation parameter $\log U$, these control nebular emission through \textsc{MAPPINGS-III} tables \citep{Levesque2010}. Dust attenuation follows the two-phase model of \cite{Charlot2000}, with $\tau_\mathrm{screen}$ for stars older than $10\,\mathrm{Myr}$ and $\tau_\mathrm{birth}$ for younger stars. Dust emission is added using \cite{Dale2014} templates, parameterised by $\alpha_\mathrm{SF,screen}$ and $\alpha_\mathrm{SF,birth}$. 

These ten physical parameters plus redshift $z$ constitute the inputs to the network. We generate training data by sampling $10^7$ galaxies in the redshift range $0<z<5$. 
 with Latin hypercube sampling. \textsc{ProSpect} is then used to compute the true observer- and rest-frame magnitudes in the five HSC bands. Sampling ranges for the input quantities are listed in Table~\ref{prior-table}, with output magnitudes spanning a wide range in $i$-band observer-frame, $5<i<35$.

\section{Network architecture and training phase}

\textsc{ProMage} is implemented in \textsc{PyTorch} \citep{Pytorch2019} as a feed-forward neural network with five hidden dense layers of $[512, 256, 128, 64, 32]$ neurons. Inputs are 11-dimensional, scaled with a standard scaler, and mapped to a single magnitude output. The dataset is split into $80\%$ training, $10\%$ validation, and $10\%$ testing. We also tested a \textsc{Speculator}-like architecture with four 128-neuron layers, but the deeper network achieved higher accuracy; an additional final $32$-neuron layer is included to stabilise extreme predictions. 

We adopt the activation function of \cite{Alsing2020}, with $\gamma$ and $\beta$ initialised to 1 and 0.1, respectively, and optimised during training. The loss function is the mean squared error, evaluated on the original magnitudes values rather than the scaled ones. Training uses the Adam optimiser with batch size $64$, a maximum of $200$ epochs, and early stopping after $20$ epochs without validation loss improvement. The learning rate starts at $10^{-3}$ and is reduced adaptively with \texttt{ReduceLROnPlateau}. Separate networks are trained for each band (observer- and rest-frame).  The network training phase is performed on the EL-9 cluster of the Leibniz Supercomputing Centre with a single Nvidia A100 GPU.

\begin{figure}
\centering

\includegraphics[width=0.49\textwidth]{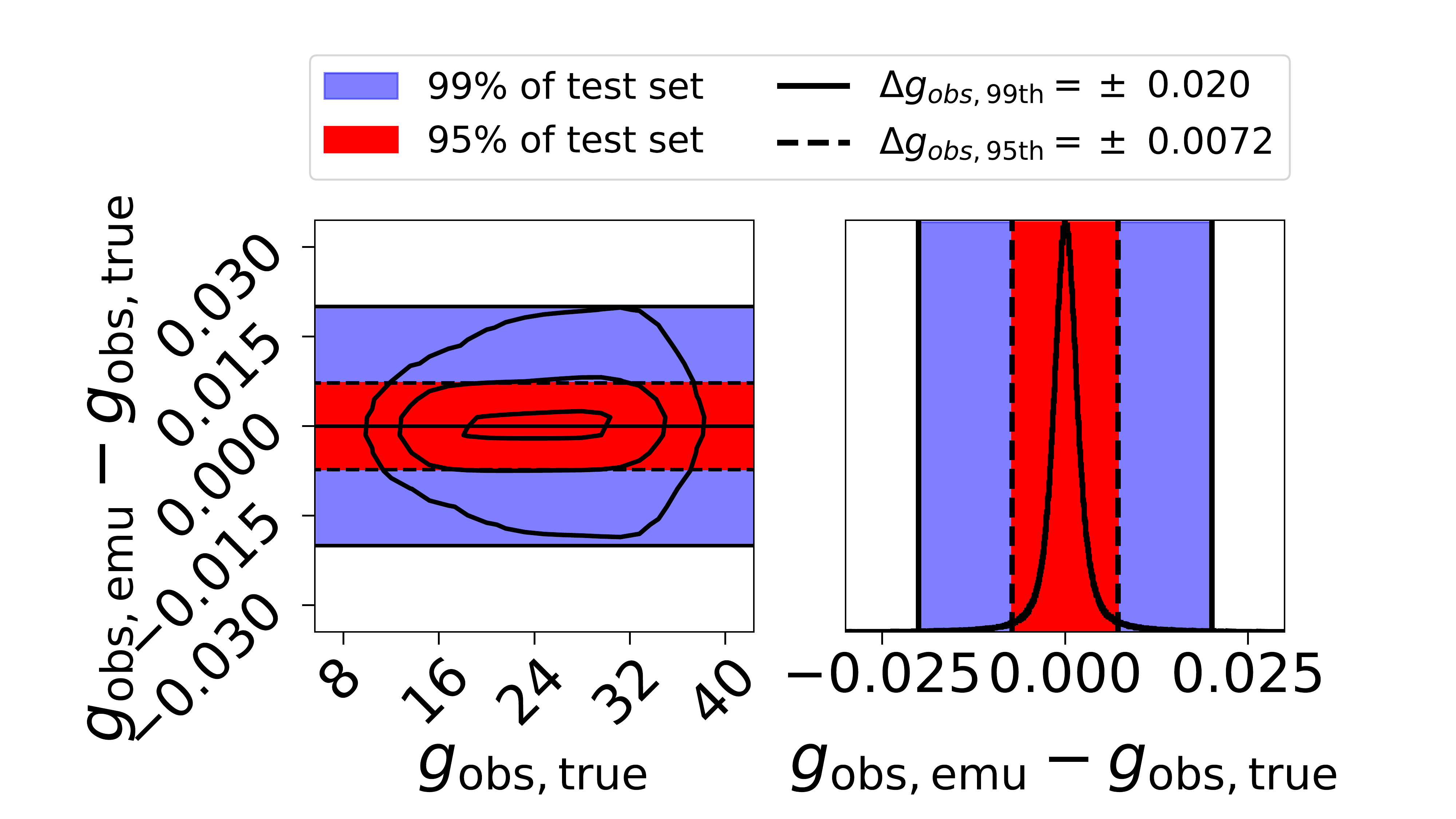} 
\includegraphics[width=0.49\textwidth]{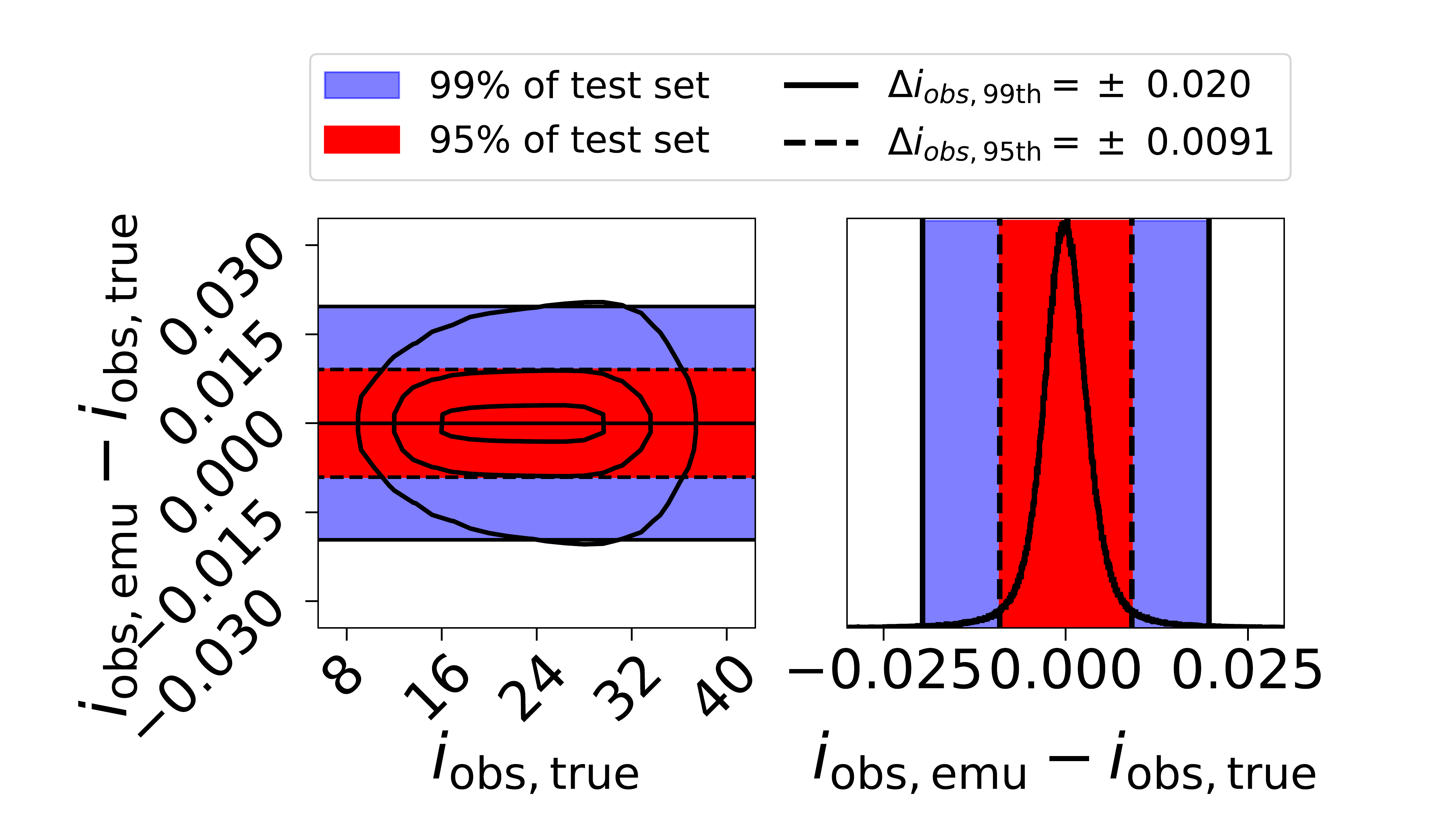}
\includegraphics[width=0.49\textwidth]{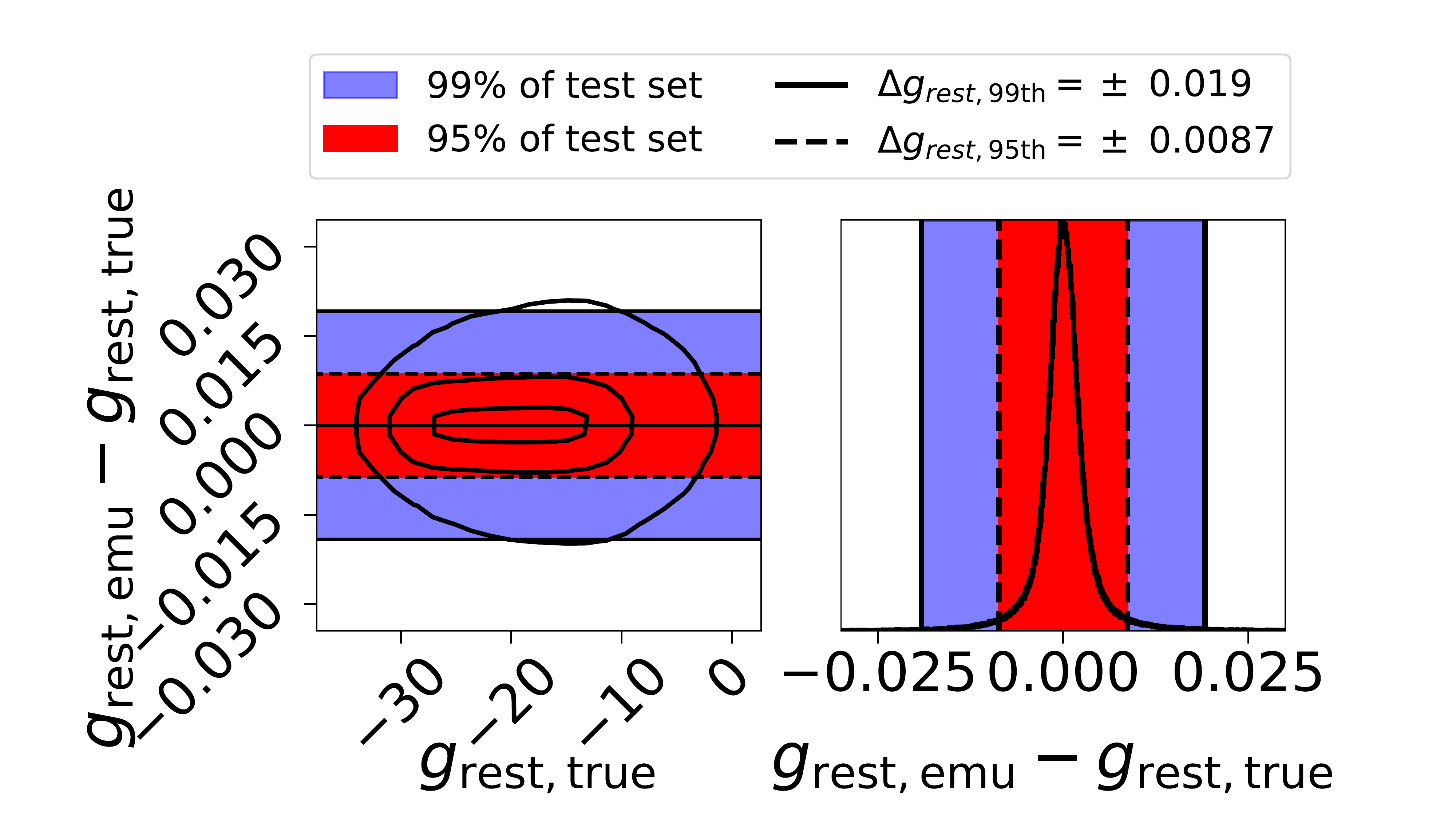} 
\includegraphics[width=0.49\textwidth]{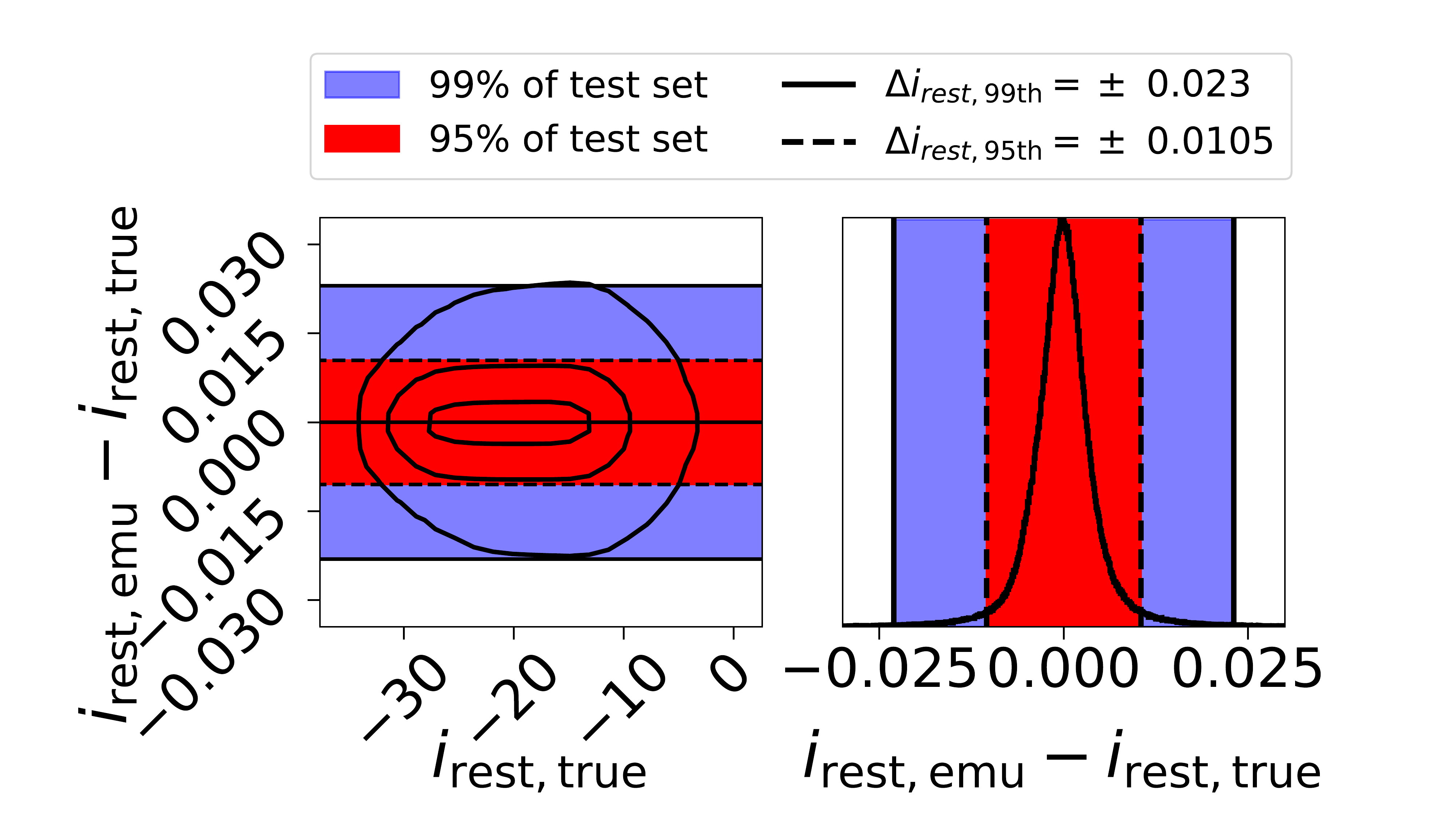}

\caption{Figure showing the performance of \textsc{ProMage} in emulating galaxy magnitudes computed with \textsc{ProSpect}.  The left panels refer to the $g$-band,  while the right panels to the $i$-band,  with upper and lower panels referring to observer (`obs') and rest-frame (`rest') magnitudes,  respectively.  We report both the histograms of the prediction accuracy and their distribution as function of the true input magnitude.  Red and blue bands represent the ranges containing $95\%$ and $99\%$ of the samples,  while the dashed and solid lines represent the $95$th and $99$th percentile values for each case.  In all reported cases, $99\%$ of the galaxies in the test set have an absolute difference with respect to the true input magnitude of $\Delta m < 0.02$. The prediction accuracy is even higher,  $\Delta m < 0.01$, for $95\%$ of galaxies.  Similar performance occurs for the other HSC optical bands.}
\label{tortorelli_figure1}
\end{figure}

\section{Results}

We show in Fig.~\ref{tortorelli_figure1} results for the $g$ and $i$ observer- and rest-frame HSC bands, representative of the overall performance. The left panels display the prediction accuracies (difference between emulated and true magnitudes) for the observer and rest-frame $g$-band, while the right panels for the $i$-band. The results are based on a test set of $10^6$ sources, evaluated in under three seconds on a Mac M1 CPU, corresponding to a $10^4$ speed-up compared to \textsc{ProSpect}. The network achieves sub-percent relative accuracy for all sources and per-mille accuracy for $99\%$ of them. Absolute errors are $< 0.02 \ \mathrm{mag}$ for $99\%$ of galaxies and $< 0.01 \ \mathrm{mag}$ for $95\%$ of them across all bands, in both observer and rest-frame. These values are below the typical photometric zero-point uncertainties of Stage~III surveys \citep{Wright2024}, and comparable or lower than the expected photometric precision of Stage IV surveys, such as Rubin-LSST \citep{Crenshaw2024}. This demonstrates that the emulator is well suited both for galaxy property inference and for forward-modelling applications, where sensitivity to changes in SED model prescriptions depends critically on photometric accuracy \citep{Tortorelli2024}.

\textit{Acknowledgements:} This work was funded by the Deutsche Forschungsgemeinschaft (DFG, German Research Foundation) under Germany’s Excellence Strategy – EXC-2094 – 390783311.  The authors gratefully acknowledge the computational and data resources provided by the Leibniz Supercomputing Centre (www.lrz.de). This project was supported in part by grant 200021\_192243 from the Swiss National Science Foundation. ASGR acknowledges funding by the Australian Research Council (ARC) Future Fellowship scheme (FT200100375).


\begin{thebibliography}{}
\bibitem[Albrecht et al.(2006)]{Albrecht2006} Albrecht, A., Bernstein, G., Cahn, R., et al.\ 2006, astro-ph/0609591. 
\bibitem[Euclid Collaboration et al.(2025)]{Mellier2024} Euclid Collaboration, Mellier, Y., Abdurrouf, et al.\ 2025, Astronomy \& Astrophysics, 697, A1. 
\bibitem[Ivezi{\'c} et al.(2019)]{Ivezic2019} Ivezi{\'c}, {\v{Z}}., Kahn, S.~M., Tyson, J.~A., et al.\ 2019, The Astrophysical Journal, 873, 2, 111. 
\bibitem[DESI Collaboration et al.(2022)]{DESI2022} DESI Collaboration, Abareshi, B., Aguilar, J., et al.\ 2022, The Astronomical Journal, 164, 5, 207. 
\bibitem[de Jong et al.(2019)]{Dejong2019} de Jong, R.~S., Agertz, O., Berbel, A.~A., et al.\ 2019, The Messenger, 175, 3. 
\bibitem[Newman \& Gruen(2022)]{NewmanGruen2022} Newman, J.~A. \& Gruen, D.\ 2022, Annual Review of Astronomy and Astrophysics, 60, 363. 
\bibitem[Fischbacher et al.(2025)]{Fischbacher2025} Fischbacher, S., Kacprzak, T., Tortorelli, L., et al.\ 2025, Journal of Cosmology and Astroparticle Physics, 2025, 6, 007. 
\bibitem[Tortorelli et al.(2025)]{Tortorelli2025} Tortorelli, L., Fischbacher, S., Gr{\"u}n, D., et al.\ 2025, arXiv:2505.21610. 
\bibitem[Conroy(2013)]{Conroy2013} Conroy, C.\ 2013, Annual Review of Astronomy and Astrophysics, 51, 1, 393. 
\bibitem[Conroy et al.(2009)]{Conroy2009} Conroy, C., Gunn, J.~E., \& White, M.\ 2009, The Astrophysical Journal, 699, 1, 486. 
\bibitem[Robotham et al.(2020)]{Robotham2020} Robotham, A.~S.~G., Bellstedt, S., Lagos, C. del P., et al.\ 2020, Monthly Notices of the Royal Astronomical Society, 495, 1, 905. 
\bibitem[Hearin et al.(2023)]{Hearin2023} Hearin, A.~P., Chaves-Montero, J., Alarcon, A., et al.\ 2023, Monthly Notices of the Royal Astronomical Society, 521, 2, 1741. 
\bibitem[Alsing et al.(2020)]{Alsing2020} Alsing, J., Peiris, H., Leja, J., et al.\ 2020, The Astrophysical Journal Supplement Series, 249, 1, 5. 
\bibitem[Cranmer et al.(2020)]{Cranmer2020} Cranmer, K., Brehmer, J., \& Louppe, G.\ 2020, Proceedings of the National Academy of Science, 117, 48, 30055. 
\bibitem[Thorp et al.(2025)]{Thorp2025} Thorp, S., Peiris, H.~V., Jagwani, G., et al.\ 2025, arXiv:2506.12122. 
\bibitem[Robotham \& Bellstedt(2025)]{Robotham2025} Robotham, A.~S.~G. \& Bellstedt, S.\ 2025, RAS Techniques and Instruments, 4, rzaf019. 
\bibitem[Bellstedt \& Robotham(2025)]{Bellstedt2025} Bellstedt, S. \& Robotham, A.~S.~G.\ 2025, Monthly Notices of the Royal Astronomical Society, 540, 3, 2703. 
\bibitem[Dotter(2016)]{Dotter2016} Dotter, A.\ 2016, The Astrophysical Journal Supplement Series, 222, 1, 8. 
\bibitem[Conroy et al.(2018)]{Conroy2018} Conroy, C., Villaume, A., van Dokkum, P.~G., et al.\ 2018, The Astrophysical Journal, 854, 2, 139. 
\bibitem[Chabrier(2003)]{Chabrier2003} Chabrier, G.\ 2003, The Publications of the Astronomical Society of the Pacific, 115, 809, 763. 
\bibitem[Driver et al.(2013)]{Driver2013} Driver, S.~P., Robotham, A.~S.~G., Bland-Hawthorn, J., et al.\ 2013, Monthly Notices of the Royal Astronomical Society, 430, 4, 2622. 
\bibitem[Bellstedt et al.(2021)]{Bellstedt2021} Bellstedt, S., Robotham, A.~S.~G., Driver, S.~P., et al.\ 2021, Monthly Notices of the Royal Astronomical Society,, 503, 3, 3309. 
\bibitem[Levesque et al.(2010)]{Levesque2010} Levesque, E.~M., Kewley, L.~J., \& Larson, K.~L.\ 2010, The Astronomical Journal, 139, 2, 712. 
\bibitem[Charlot \& Fall(2000)]{Charlot2000} Charlot, S. \& Fall, S.~M.\ 2000, The Astrophysical Journal, 539, 2, 718. 
\bibitem[Dale et al.(2014)]{Dale2014} Dale, D.~A., Helou, G., Magdis, G.~E., et al.\ 2014, The Astrophysical Journal, 784, 1, 83. 
\bibitem[Paszke et al.(2019)]{Pytorch2019} Paszke, A., Gross, S., Massa, F., et al.\ 2019, , arXiv:1912.01703.
\bibitem[Wright et al.(2024)]{Wright2024} Wright, A.~H., Kuijken, K., Hildebrandt, H., et al.\ 2024, Astronomy \& Astrophysics, 686, A170. 
\bibitem[Crenshaw et al.(2024)]{Crenshaw2024} Crenshaw, J.~F., Kalmbach, J.~B., Gagliano, A., et al.\ 2024, The Astronomical Journal, 168, 2, 80. 
\bibitem[Tortorelli et al.(2024)]{Tortorelli2024} Tortorelli, L., McCullough, J., \& Gruen, D.\ 2024, Astronomy \& Astrophysics, 689, A144. 


\end{thebibliography}
\end{document}